\documentclass{PoS}

\usepackage{graphicx}

\title{Blasting through lattice calculations using CUDA }

\ShortTitle{Blasting through lattice calculations using CUDA }

\author{Kipton Barros\\
        Department of Physics, Boston University,
        Boston, MA 02215\\
        E-mail: \email{kbarros@bu.edu}}
\author{Ronald Babich\\
        Department of Physics, Boston University,
        Boston, MA 02215\\
        E-mail: \email{rbabich@bu.edu}}
\author{Richard Brower\\
        Department of Physics, Boston University,
        Boston, MA 02215\\
        E-mail: \email{brower@bu.edu}}
\author{Michael A.~Clark\\
        Center for Computational Science, Boston University,
        Boston, MA 02215\\
        E-mail: \email{mikec@bu.edu}}
\author{\speaker{Claudio Rebbi}\\
        Department of Physics, Boston University,
        Boston, MA 02215\\
        E-mail: \email{rebbi@bu.edu}}

      \abstract{ Modern graphics hardware is designed for highly
        parallel numerical tasks and provides significant cost and
        performance benefits.  Graphics hardware vendors are now
        making available development tools to support general purpose
        high performance computing.  Nvidia's CUDA platform, in
        particular, offers direct access to graphics hardware through
        a programming language similar to C.  Using the CUDA platform
        we have implemented a Wilson-Dirac operator which runs at an
        effective 68 Gflops on the Tesla C870.  The recently
        released GeForce GTX 280 runs this same code at 92 Gflops, and we
        expect further improvement pending code optimization.  }

\FullConference{The XXVI International Symposium on Lattice Field Theory \\
		 July 14 - 19, 2008\\
		 Williamsburg, Virginia, USA}

\begin{document}

\section{Introduction}

For decades, Moore's law has reliably given a doubling of the number
of transistors per chip about every two years, a trend that continues
to this day.  In the past, such increases translated directly into
improved performance for serial code through higher clock rates,
larger caches, and increased exploitation of instruction-level
parallelism.  Recently, however, such improvements have yielded
diminishing returns, bringing us to the era of multi-core CPUs.  For
the intrinsically parallel tasks commonly found in scientific
computing, this is a welcome development.  Still, it is not obvious
that commodity processors, whose high clock rates and large caches
come at the expense of greater numbers of cores, represent the optimal
balance for highly parallel workloads.  Graphics processing units
(GPUs), driven by the enormous video game market, represent a
different set of trade-offs.  GPUs emphasize very high parallelism and
memory bandwidth, a recipe for astounding performance in many
scientific applications.

In lattice gauge theory (LGT), the application of the Wilson-Dirac
operator is a performance critical task and has been shown to map well
onto GPU architectures~\cite{Egri:2006zm,ibrahim-gpu}.  We describe an
implementation of the Wilson-Dirac operator which runs at an effective
92 Gflops on the Nvidia GTX 280 GPU. Our implementation uses Nvidia's
CUDA (Compute Unified Device Architecture) platform.

CUDA is a C-like programming language and a full software development
toolkit~\cite{cudaguide}.  CUDA provides direct and relatively
low-level access to the GPU. Current high-end Nvidia graphics cards are
specifically designed to facilitate general purpose computation
through CUDA. An important feature, which is available in CUDA but not
previous frameworks, is high speed synchronization and data sharing between
threads.

Besides CUDA, several other platforms exist for highly threaded
computation. Advanced Micro Devices (AMD) is another leading vendor of
GPUs and has introduced the Stream SDK toolkit~\cite{stream}. The Cell
processor, originally designed by Sony, Toshiba, and IBM for consumer
applications, is also an attractive target for
LGT~\cite{Belletti:2007pp,Spray:2008nt,Baier:2008kv}. Even among
mainstream CPUs there is a trend toward increasing numbers of parallel
cores. For example, Intel's Larrabee architecture is expected to have
tens of cores, blurring the distinction between CPU and GPU
architectures~\cite{larrabee}.  A potential advantage of Nvidia's CUDA
platform is that it may be used to target both GPUs and multi-core
CPUs.  Finally, we note that the proposed OpenCL
standard~\cite{opencl} describes a programming model quite similar to
CUDA.

\section{Hardware}

Nvidia produces three lines of graphics cards. The GeForce series
serves the lucrative consumer video game market, while the Tesla
series targets the high performance computing (HPC) market. Tesla
cards retain the core architecture of the consumer cards but offer
more device memory and greater reliability, at the expense of lower
memory bandwidth and increased cost.\footnote{The Tesla series also
  lacks video output.}  Finally, Nvidia markets the Quadro line for
professional graphics applications.

\begin{table}
\begin{center}
    \begin{tabular}{ | l | l | l | l | l | l | }
      \hline
      Card & Cores & Bandwidth (GB/s) & Gflops & Device Memory \\ \hline
      GeForce 8800 GTX & 128 & 86.4 & 518 & 768 MB \\
      Tesla C870 & 128 & 76.8 & 518 & 1.5 GB \\
      GeForce GTX 280 & 240 & 141.7 & 933 & 1 GB \\
      Tesla C1060 & 240 & 102 & 933 & 4 GB \\ \hline
    \end{tabular}
    \caption{\label{specs}Specifications of representative Nvidia
      cards.  The Tesla S1070 (not listed) is a 1U unit, containing
      the equivalent of four Tesla C1060 cards.}
\end{center}
\end{table}

To date, there have been roughly two generations of CUDA-enabled GPUs.
The flagship consumer cards of the previous and current generation are
the GeForce 8800 GTX and the GTX 280, paralleled in the HPC line by
the Tesla C870 and C1060 (variants are also available in a single
rack-mountable unit containing multiple GPUs).  See Table~\ref{specs}
for detailed specifications.  In Section~\ref{perf} below, we
benchmark our code on both the Tesla C870 and GeForce GTX 280.  The
Tesla C1060 was not available for benchmarking at the time of the
conference.

\begin{figure}
\begin{center}
\includegraphics*[width=0.6\textwidth]{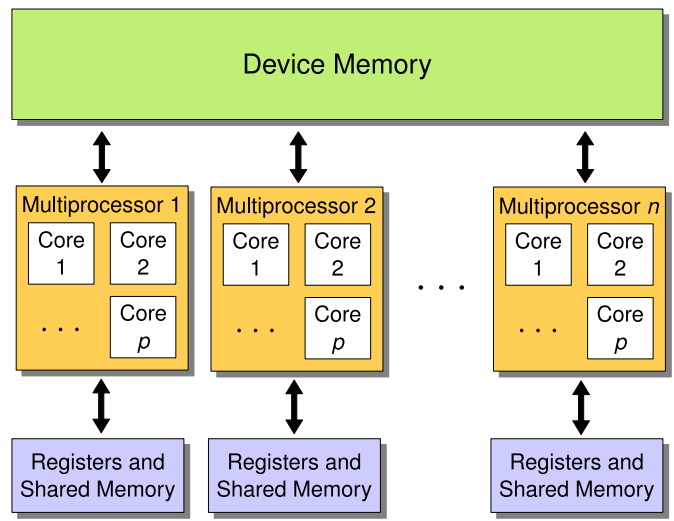} 
\caption{\label{fig-arch}Architecture of a modern Nvidia graphics card.
In Nvidia's nomenclature, cores are called \emph{stream processors}
(or \emph{scalar processors}), and in current GPUs each multiprocessor
has eight such cores.}
\end{center}
\end{figure}

A modern Nvidia GPU contains many multiprocessors, each composed of
several cores, as illustrated in Fig.~\ref{fig-arch}.  For example,
the GPU in the Tesla C1060 contains 30 multiprocessors and a total of
240 cores.  Within a multiprocessor, cores are allocated local
registers and have access to a fast shared memory. In addition,
each multiprocessor provides two small read-only caches: a constant
cache, and a texture cache to speed up global device reads.  Primary
storage on the card is provided by \emph{device memory}, which is
shared among all multiprocessors and has a relatively high
latency. However, this latency can often be hidden by having a large
number of threads ready to execute.  A very important
consideration is that the highest bandwidth from device memory is
achieved when accesses are coalesced; this occurs when groups of 16
threads access a contiguous, properly aligned memory region.\footnote{This
  requirement has been relaxed somewhat in the more recent generation
  of cards.}

\section{The CUDA programming model}

The CUDA platform provides direct access to the GPU through a C-like
programming language with minimal extensions. The CUDA platform
includes a compiler that targets the GPU, as well as a hardware driver
and a runtime library.  Higher level libraries are also provided,
including optimized BLAS and FFT implementations.

A CUDA application works by spawning a very large number of threads,
as many as tens of thousands at once, which execute in parallel.  For
example, in a LGT application one might assign one thread to each
lattice site.  The user specifies the number, organization, and shared
memory usage of the threads when a CUDA kernel is invoked.  As an
example, consider the CUDA code
\begin{verbatim}
        dslashKernel <<<gridDim, blockDim, sharedBytes>>> (args);
\end{verbatim}
which invokes {\tt dslashKernel(args)} for execution by many
individual threads on the GPU. Threads are grouped into \emph{thread
  blocks,} and the entire collection of thread blocks is called a
\emph{grid}.  The code above tells the GPU to launch a kernel using
{\tt gridDim} blocks, each containing {\tt blockDim} threads.
The compiler is instructed to allocate {\tt sharedBytes} bytes of
\emph{shared memory} to each block. This shared memory allows for
rapid communication between threads within a thread block.  CUDA
provides primitives to allow synchronization between threads
within a thread block. However, no synchronization is possible
between different thread blocks (within a single kernel invocation).

The GPU will dynamically schedule the thread blocks for execution. In
order to hide high latency operations, it is desirable to have a high
\emph{multiprocessor occupancy:} each multiprocessor should have many
threads simultaneously loaded and waiting for execution. The
challenges to achieving high multiprocessor occupancy will be
discussed in Section~\ref{localmem}. The GPU supports conditional
execution, but it is highly desirable that groups of 32 threads (a
\emph{thread warp}) follow the same execution path. Otherwise, both
execution paths are serialized and executed by the entire warp.

\section{Wilson inverter: Main features}

We implemented a CUDA kernel, Dslash, which calculates the
Wilson-Dirac operator restricted to spinor variables at the even or
odd parity sites. We have also implemented an even-odd
preconditioned conjugate gradient (CG) solver.

The action of Dslash and Dslash$^\dagger$ on a spinor field is the
most compute-intensive part of the CG inverter. Our CUDA
implementation spawns one thread for each site in the (even or odd)
sub-lattice.  Performance is limited by memory bandwidth, as is common
in LGT applications.  Nonetheless, the GPU achieves a high sustained
performance due to the relatively large memory bandwidth.  We have
structured the calculation in a manner which helps reduce the amount
of transferred data (e.g., we mostly use 32-bit precision, we gauge
fix to the temporal gauge, we reconstruct the gauge links from only
two columns of data, etc.) and maximizes the overlap of computation
and communication.

The CG algorithm is decomposed into several operations, each of which
runs as a separate CUDA kernel. Besides the Dslash operator, we also
require a dot product operation and a variety of BLAS-like
operations. The dot product involves a global sum, which is performed
by the GPU in double precision using parallel reduction. In the
Wilson-Dirac solver, the largest portion of compute time is spent in
applying the Dslash and Dslash$^\dagger$ operators. We find that the
whole solver runs at well over 80 percent of the speed of just its
Dslash component.  We also emphasize that the operation of the
Wilson-Dirac solver involves negligible data transfer between the
device and host computer.

\section{Data layout}

We consider a lattice of $3+1$ dimensions, splitting the gauge and
spinor fields into even and odd sub-lattices.  With three colors and
four spin components, spinor fields require 24 floats per lattice site
to store.  Gauge fields require 18 floats per link.
Following~\cite{Egri:2006zm}, we employ a specialized data layout. We
do so because, as we have mentioned, maximum bandwidth is obtained
when 16 consecutive threads (a \emph{half warp}) simultaneously read
16 primitive elements which are packed contiguously in device
memory. The available primitive elements include structures of 1, 2,
or 4 packed floats. Our testing indicates that, on current hardware,
the best bandwidth is achieved using device reads through the texture
cache, and using {\tt float4} primitives. Spinor objects are composed
of 24 floats, so we use 6 arrays of {\tt float4}s to store the entire
spinor field. In this way, consecutive threads can simultaneously
access (nearly) consecutive elements from device memory. The gauge
links are stored in 12 floats (before SU(3) reconstruction), requiring
3 arrays of {\tt float4}s.

Because the lattice sites are split by parity, and because of boundary
effects, the half warp of 16 consecutive threads may access {\tt
  float4} objects which are nearly, but not exactly, contiguous in
memory. The texture cache mitigates the performance penalty of
imperfect memory accesses.

\section{\label{localmem}Local storage constraints}

Unlike CPUs, the GPU does not provide a large memory cache. Instead,
the GPU provides a relatively small amount of fast \emph{shared
  memory,} which is manually managed and shared between
threads. Shared memory is orders of magnitude faster than device
memory, so access to the latter must be minimized. For the Dslash
operation, we have found that shared memory alone is not satisfactory
for local data storage, and we therefore employ registers for data
storage as well.

For hiding high latency operations it is important to have a high
multiprocessor occupancy: many \emph{active threads} should be
simultaneously loaded and ready to execute. Our tests indicate that
192 active threads per multiprocessor is desirable and that
performance rapidly degrades with fewer active threads. The
multiprocessor occupancy is determined by the register and shared
memory requirements of the CUDA kernel.

At the hardware level, a single GPU contains many multiprocessors
(currently 16 or 30). Each multiprocessor has a fixed quantity of
local storage (registers and shared memory) which is divided among
many active threads. Nvidia's previous generation GPUs provide 8,192
registers and 16 KB of shared memory per multiprocessor, while in the
more recent generation the number of registers has been increased to
16,384.

What are the local storage requirements for each thread in the Dslash
operation? Each thread must accumulate to a single output spinor,
which is composed of 24 floats and should reside in local storage. In
constructing the output spinor, the thread loops over all neighboring
sites. In each direction, a full spinor and gauge link must be
read. The neighboring spinor is immediately projected into a
\emph{half spinor} and requires only 12 floats of local storage. The
SU(3) matrix representing the gauge link requires an additional 18
floats. Thus, at a minimum, 54 floats are required per thread. If
these are to be stored entirely in the 16 KB of shared memory, then at
most 64 threads would be active on one multiprocessor.\footnote{The
  number of active threads must be a multiple of 32, the warp size. A
  multiple of 64 is recommended.}  This number is much smaller than
the target, 192, and would negatively impact performance. Our trick is
to use registers for additional data storage. The GPU has at least
8,192 registers per multiprocessor, providing 32 KB of local
storage. Using both shared memory and registers it is possible to
obtain 192 active threads per multiprocessor for the Dslash kernel.

What are registers exactly?  Registers serve the same function on a
GPU as they do on a CPU. Namely, they appear as explicitly labeled
operands in machine code instructions generated by the compiler. Every
active thread on a multiprocessor is allocated a sufficient number of
private registers to execute the CUDA kernel. Unlike shared memory,
registers cannot be shared between threads. Another limitation is that
data stored in registers cannot be organized into an array and
dynamically indexed. For example, we store the SU(3) matrix elements
in registers by declaring
\begin{verbatim}
float g1, g2, g3, ..., g18;
\end{verbatim}
We cannot use loops to express matrix operations on these
elements. Writing the full Dslash operation by hand, and without using
loops, would be tedious and error-prone. For this reason, we found it
expedient to automatically generate the lengthy Dslash CUDA code. The
Dslash code generator was written in the Scala programming language.

\section{\label{perf}Performance}

\begin{figure}
\begin{minipage}[t]{0.49\textwidth}
\includegraphics*[width=\textwidth]{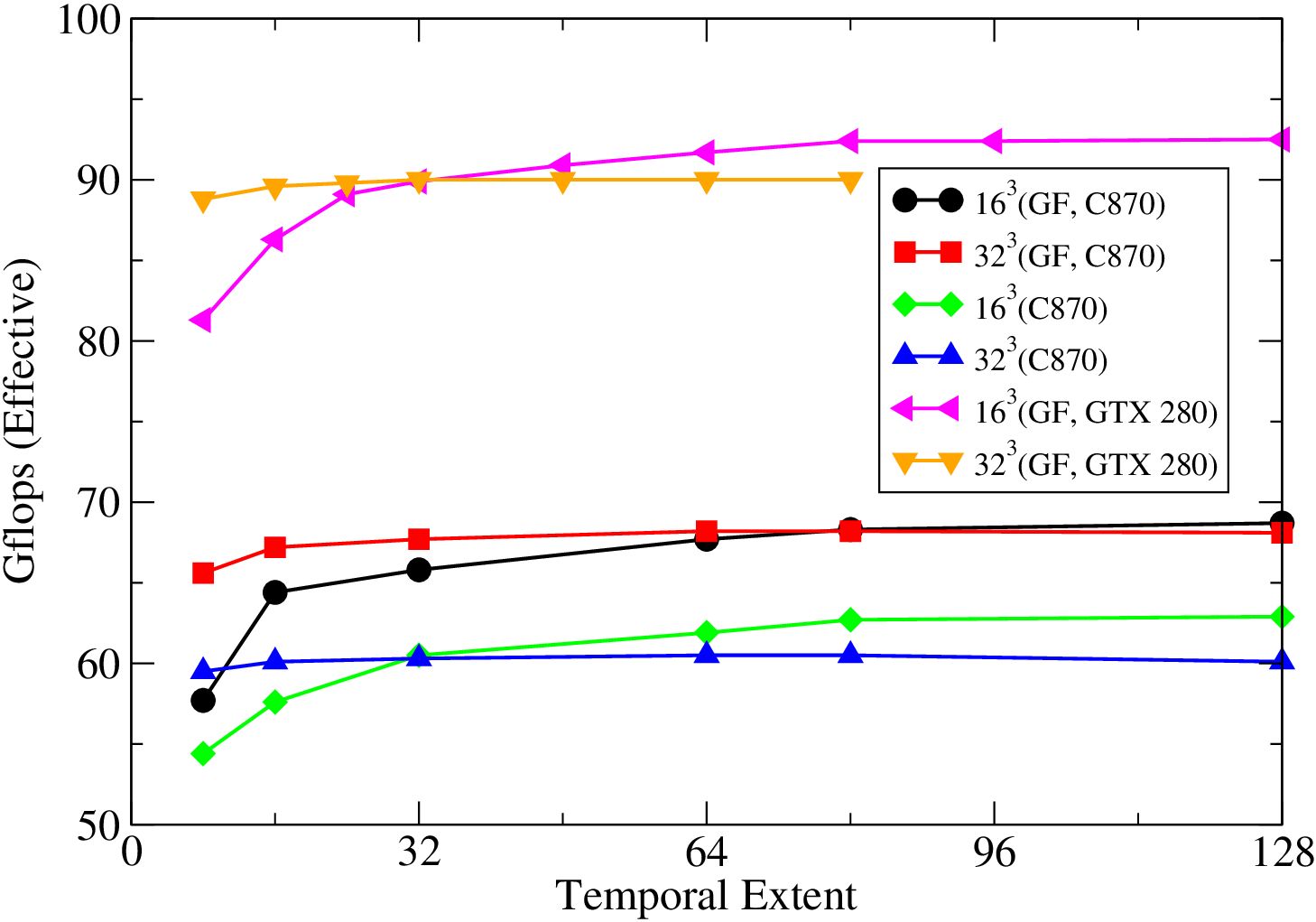} 
\caption{\label{fig-dslash}Performance of the preconditioned Wilson operator
  $(1-\kappa^2 D_{eo} D_{oe})$.}
\end{minipage}
\hfill
\begin{minipage}[t]{0.49\textwidth}
\includegraphics*[width=\textwidth]{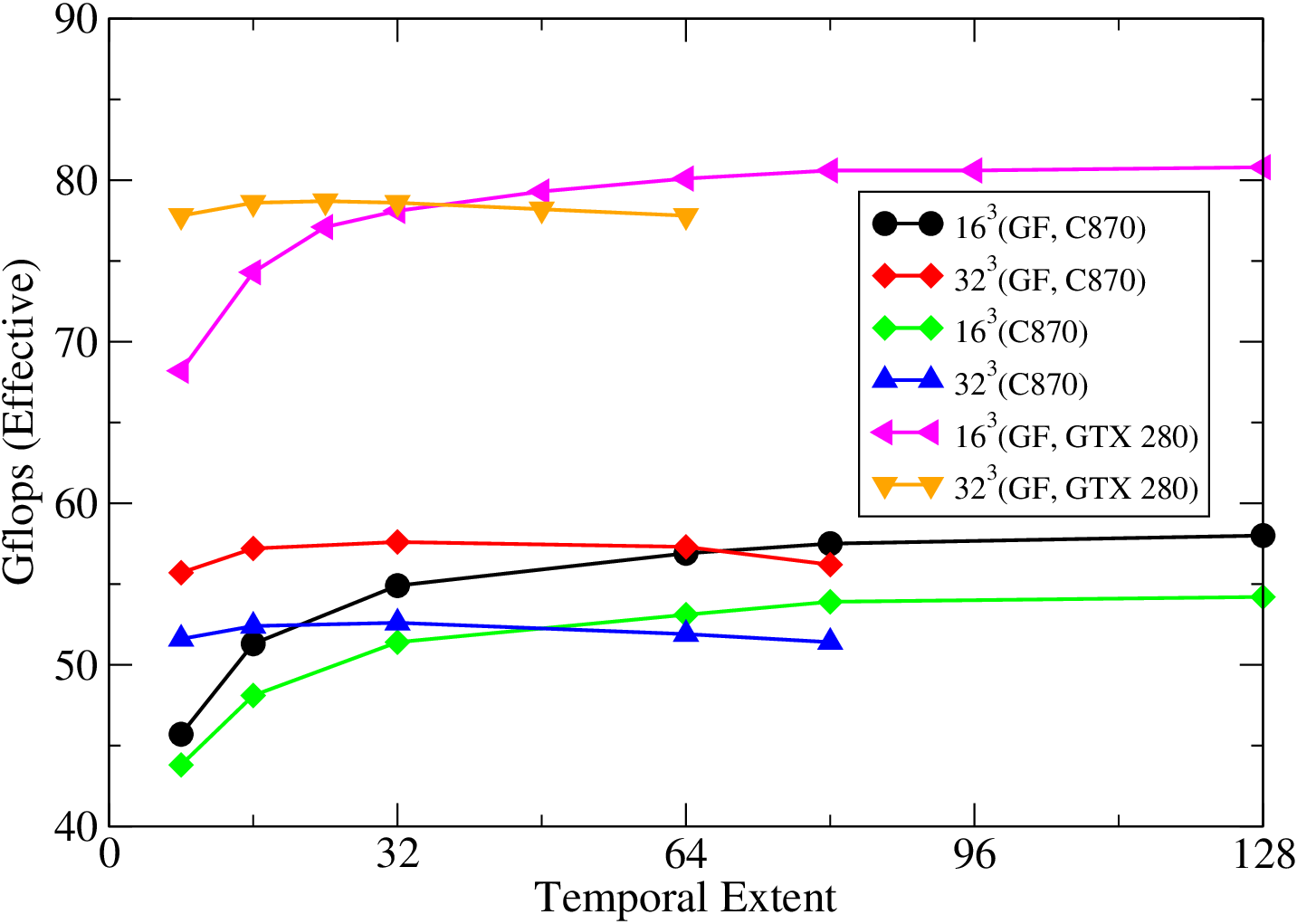} 
\caption{\label{fig-cg}Performance of the conjugate gradient inverter.}
\end{minipage}
\end{figure}

In Fig.~\ref{fig-dslash}, we present performance results for the
even-odd preconditioned Wilson-Dirac operator on a range of different
volumes.  We find that the performance is only weakly
volume-dependent: for all but the smallest volumes it sustains above
60~GFlops on the C870 and around 90~GFlops on the GTX 280.  Results
for the full CG inverter are shown in Fig.~\ref{fig-cg}.  For volumes
of reasonable size, the inverter sustains over 50~GFlops and 80~GFlops
on the C870 and GTX 280, respectively.  We have also implemented a
BiCGstab inverter that achieves similar performance.

The (GF) label on the plots signify that the gauge-fixing trick was
used.  This involves fixing the temporal gauge links to the identity
(on all but one time-slice) in order to save memory bandwidth, and
improves performance by about 10 percent.  We also save bandwidth by
reading only two columns of each gauge matrix and reconstructing the
third on the fly.  In all cases, the reported performance numbers are
``effective gigaflops'' that may be compared with implementations on
traditional architectures.  In particular, the nominal number of
operations per lattice site does not include the extra work done in
the SU(3) reconstruction, nor the savings associated with having
trivial links in the time direction.

We note that performance of the GPU code is more than an order of
magnitude greater than typical SSE-optimized implementations (which
generally achieve less than 5~Gflops for Wilson matrix-vector on a
3.0~GHz quad-core Xeon processor).  In addition, the scaling of
performance with volume is relatively constant, as compared to CPU
implementations which suffer dramatically as the local volume falls
out of cache.  At 80 Gflops sustained CG performance and \$450 per
board, the GTX 280 card represents a price-performance ratio of
\$5.60/Gflops (excluding, of course, the cost of the host computer).

\begin{acknowledgments}
This work was supported in part by US DOE
grants DE-FG02-91ER40676 and DE-FC02-06ER41440 and NSF grants DGE-0221680,
PHY-0427646, and OCI-0749300.
\end{acknowledgments}

\end{document}